\documentclass[prl,twocolumn,floatfix,a4paper,superscriptaddress]{revtex4}
\usepackage{bm,color,graphicx,amsmath,txfonts}

\newcommand{\GG}{{\cal G}}

\newcommand{\NN}{{\cal N}}
\newcommand{\PP}{{\cal P}}
\newcommand{\QQ}{{\cal Q}}

\begin{document}

\title{Generation and detection of non-Gaussian phonon-added coherent states in optomechanical systems}

\author{Jie Li}
\affiliation{Department of Physics, Zhejiang University, Hangzhou 310027, China}
\affiliation{Institute for Quantum Science and Engineering and Department of Biological and Agricultural Engineering, Texas A{\rm \&}M University, College Station, Texas 77843, USA}
\author{Simon Gr\"oblacher}
\affiliation{Kavli Institute of Nanoscience, Delft University of Technology, 2628CJ Delft, Netherlands}
\author{Shi-Yao Zhu}
\affiliation{Department of Physics, Zhejiang University, Hangzhou 310027, China}
\author{G. S. Agarwal}
\affiliation{Institute for Quantum Science and Engineering and Department of Biological and Agricultural Engineering, Texas A{\rm \&}M University, College Station, Texas 77843, USA}
\affiliation{Department of Physics and Astronomy, Texas A{\rm \&}M University, College Station, Texas 77843, USA}

\begin{abstract}
Adding excitations on a coherent state provides an effective way to observe nonclassical properties of radiation fields. Here we describe and analyse how to apply this concept to the motional state of a mechanical oscillator and present a full scheme to prepare non-Gaussian {\it phonon}-added coherent states of the mechanical motion in cavity optomechanics. We first generate a mechanical coherent state using electromagnetically induced transparency. We then add a single phonon onto the coherent state via optomechanical parametric down-conversion combined with single photon detection. We validate this single-phonon-added coherent state by using a red-detuned beam and reading out the state of the optical output field. This approach allows us to verify nonclassical properties of the phonon state, such as sub-Poissonian character and quadrature squeezing. We further show that our scheme can be directly implemented using existing devices, and is generic in nature and hence applicable to a variety of systems in opto- and electromechanics.

\end{abstract}

\date{\today}
\maketitle

The realization of nonclassical states of light has opened up the possibility of using quantum optics for a variety of applications in quantum sensing~\cite{GAbook,Bowen,ZYOu}, and quantum information science~\cite{qiRMP1,qiRMP2}, as well as for extremely sensitive measurements~\cite{qubitRMP}. Optomechanical systems on the other hand offer an exciting opportunity to study quantum states of macroscopic systems~\cite{omRMP}. Considerable progress has been made in controlling these massive systems down to the quantum level. Recent breakthroughs include optomechanical squeezing of light~\cite{sqzO1,sqzO2} and of mechanical motion~\cite{sqzM1,sqzM2}, quantum entanglement between mechanics and a cavity field~\cite{enOM1,enOM2}, as well as between two mechanical oscillators~\cite{enMM1,enMM2}, and nonclassical photon-phonon correlations by measuring the second-order cross-correlation $g^{(2)}$~\cite{g2}.

Recent interest has focused on generating non-Gaussian states of mechanical systems, e.g., preparing the mechanical oscillator in a single-phonon Fock state. Following several theory proposals~\cite{enOM2,Galland}, this has been realized by exploiting the optomechanical parametric down-conversion combined with single photon detection~\cite{Simon17} in close analogy to early quantum optics experiments. It could also be generated by transferring the single-photon state from an optical field to a mechanical resonator (MR)~\cite{Chen}. Alternative ways to create non-Gaussian states of a mechanical system include exploiting the intrinsic nonlinearity of the optomechanical interaction~\cite{Girvin}, or by making measurements on the optical field~\cite{Mauro}. The preparation of such nonclassical states of a massive object is important in connection with the studies of quantum effects at the macroscopic scale~\cite{macroQM,CMs}. There are many other non-Gaussian states, such as quantum superposition states~\cite{superX}, and excitation-added/subtracted coherent and squeezed states~\cite{GAgrc,GA91,GA07}. The addition of excitations on a coherent state, for example, provides a way to observe quantum effects of a radiation field, such as quadrature squeezing, sub-Poissonian character, and negative Wigner distributions~\cite{GA91}. Single-photon-added coherent states of light have been generated using parametric down-conversion in a nonlinear crystal in combination with single photon detection~\cite{Bellini04}. In this paper, we apply this concept to the motional state of a mechanical oscillator. Specifically, we work on cavity optomechanics~\cite{omRMP} and provide a full scheme to generate and detect {\it single-phonon}-added coherent states (PACS) of mechanical motion~\cite{GAgrc}. As a first step, we prepare the MR in a coherent state using electromagnetically induced transparency~\cite{GAEIT}, or equivalently optomechanically induced transparency (OMIT)~\cite{OMIT}, where the cavity is bichromatically driven by a strong red-detuned field and a much weaker field on cavity resonance. The former can also help cool the mechanical motion close to its quantum ground state~\cite{cool1,cool2}, and the latter is used to displace the ground state in phase space to a coherent state. We then add a single phonon onto the coherent state via optomechanical parametric down-conversion combined with single photon detection. This can be realized by sending a weak blue-detuned laser pulse into the optomechanical cavity. Finally we confirm the generated phonon state by using a relatively strong red-detuned pulse which realizes a state swap operation between the MR and the light pulse. We then measure the Mandel $Q$ parameter of the cavity output field, confirming the sub-Poissonian character of the mechanical state. Alternatively, homodyning the cavity output and measuring the variance of the quadrature can be used to detect squeezing of the mechanical state. Lastly, we analyse the effects of residual thermal excitations in the coherent-state-preparation stage on the results of the ideal case where the mechanical motion is cooled exactly into its quantum ground state (mean phonon number $\bar{n}_0=0$).

\begin{figure}[b]
\hskip0cm\includegraphics[width=0.95\linewidth]{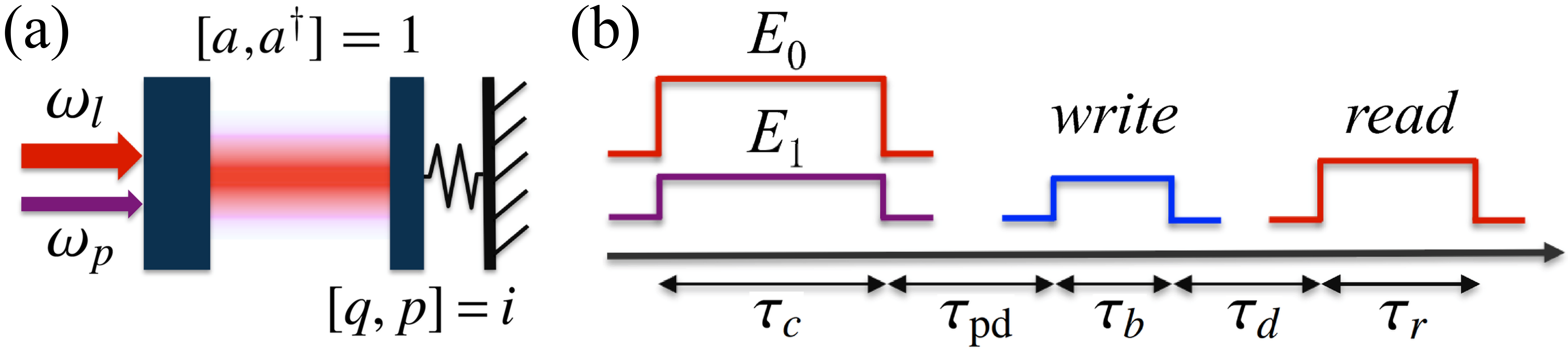} 
\caption{(a) Sketch of the generic system for preparing coherent states of the mechanical motion. (b) Pulse sequence of the scheme. Two laser pulses of duration $\tau_c$ are used to generate a mechanical coherent state. Once the MR is prepared in a desired state, the two lasers are switched off and after some time $\kappa^{-1} \ll \tau_{\rm pd}\ll \gamma^{-1}$, during which all cavity photons decay, while the mechanical state remains unchanged, a write pulse is sent. This blue-detuned pulse of duration $\tau_b$ prepares the MR in a single-PACS provided that a single photon is detected in the interval $\tau_d$. The red-detuned readout pulse of duration $\tau_r$ then transfers the mechanical state to the cavity output field for subsequent measurements. In order to neglect mechanical damping, $\tau_{\rm pd}+\tau_b+\tau_d+\tau_r \ll \gamma^{-1}$ is assumed. }  
\label{fig1}
\end{figure}

{\it Preparing mechanical coherent states.} We consider an optical cavity mode with resonance frequency $\omega_c$ and annihilation (creation) operator $a$ ($a^{\dag}$) ($[a,a^{\dag}]\,{=}\,1$) coupled to a MR with frequencies $\omega_m$ via radiation pressure, as depicted in Fig.~\ref{fig1} (a). The cavity mode is bichromatically driven by a strong red-detuned field at $\omega_l \,{\simeq}\, \omega_c \,{-}\, \omega_m$ and a much weaker field on cavity resonance $\omega_p \, {\simeq} \, \omega_c$. The Hamiltonian of the system reads
\begin{equation}
\begin{split}
H&=\hbar \omega_c a^{\dag} a +\frac{1}{2} \hbar\omega_m (q^2+p^2) -\hbar g a^{\dag} a q  \\
&+i \hbar [( E_0 e^{-i \omega_l t} + E_1 e^{-i \omega_p t}) a^{\dag} - {\rm h.c.} ].
\end{split}
\end{equation}
Here, $q$ and $p$ are the dimensionless position and momentum quadratures of the MR satisfying the commutation relation $[q,p]\,{=}\,i$. $g$ is the single-photon optomechanical coupling rate. $E_0$ and $E_1$ are respectively related to the power of the driving fields $P_0$ and $P_1$ ($P_1 \ll P_0$) by $|E_0|^2=2\kappa P_0/\hbar \omega_l$ and $|E_1|^2=2\kappa P_1/\hbar \omega_p$, where $\kappa$ is the cavity decay rate. Such a system has been employed for the study of OMIT in cavity optomechanics~\cite{GAEIT,OMIT}. Instead of looking at the {\it mean field} response of the system to the probe field, we focus on the mechanical part here and include quantum fluctuations.  The mechanical motion can be cooled close to its quantum ground state ($\bar{n}_0 \ll 1$) provided that the system is working in the resolved sideband limit and has a large cooperativity~\cite{cool1,cool2}. We work in this deep cooling regime and show that by including a weak ``probe" field, the MR can be prepared in a coherent state, where the averages of the mechanical quadratures show periodic behaviors~\cite{Mari}, while the fluctuations almost remain unchanged. This can easily be understood: the red-detuned strong light beam realizes an effective beamsplitter interaction which maps the coherent state of the weak field on cavity resonance onto the mechanical state~\cite{Wang}.

By taking average values and using the factorization $\langle AB \rangle \simeq \langle A \rangle \langle B \rangle$ ($A$ and $B$ are arbitrary system operators), we obtain the Langevin equations responsible for the first moments, which in the reference frame rotating at $\omega_l$ are
\begin{equation}
\begin{split}
\langle \dot{q} \rangle&=\omega_m \langle p \rangle, \,\,\,\,\,\,\,\,
 \langle \dot{p} \rangle= -\omega_m \langle q \rangle - \gamma \langle p \rangle +g \langle a^{\dag} \rangle  \langle a \rangle ,  \\
\langle \dot{a} \rangle&=-(\kappa + i \Delta_0) \langle a \rangle + i g \langle a \rangle \langle q \rangle +E_0 + E_1 e^{-i \delta t},  \\
\end{split}
\label{Aveq}
\end{equation}
where $\gamma$ denotes the mechanical damping rate, and $\Delta_0{=}\omega_c{-}\omega_l$ and $\delta{=}\omega_p {-} \omega_l$. In the long time limit, $t \gg \gamma^{-1}$, all average values have the form $\langle O \rangle\,{=}\,\sum_{n=-\infty}^{+\infty} e^{-i n \delta t} \langle O \rangle_n$ ($O\,{=}\,q,p,a$)~\cite{GAbook}, where $\langle O \rangle_n$ are time independent. The substitution of $\langle O \rangle$ in Eq.~\eqref{Aveq} yields an hierarchy of coupled equations. Nevertheless, by assuming a weak ``probe" field $|E_1| \,{\ll}\, |E_0|$, one can terminate the series in $\langle O \rangle$ at $n\,{=}\,1$. Substituting the truncated series of $\langle O \rangle$ in Eq.~\eqref{Aveq} and equating the coefficients of different Fourier components, simple approximated solutions of the averages can be obtained~\cite{SuppM}: $\langle a \rangle \simeq \langle a \rangle_0 + \langle a \rangle_1 e^{-i \omega_m t} $ for the cavity field, where $\langle a \rangle_0 \simeq \frac{E_0}{\kappa+i \omega_m}$, $\langle a \rangle_1 \simeq \frac{E_1}{\kappa+\frac{g^2}{\gamma}|\langle a \rangle_0|^2}$, and
\begin{equation}\label{qpSolu}
\begin{split}
\langle q \rangle &= \langle q \rangle_0 + 2 {\rm Re}\langle q \rangle_1 \cos \omega_m t + 2 {\rm Im}\langle q \rangle_1 \sin \omega_m t,  \\
\langle p \rangle &=  2 {\rm Re}\langle p \rangle_1 \cos \omega_m t + 2 {\rm Im}\langle p \rangle_1 \sin \omega_m t
\end{split}
\end{equation}
for the mechanical mode, $\langle q \rangle_0 \simeq \frac{g}{\omega_m} (|\langle a \rangle_0|^2{+}|\langle a \rangle_1|^2)$, $\langle q \rangle_1  \simeq  \frac{i g}{\gamma} \langle a \rangle_0^* \langle a \rangle_1$, and $\langle p \rangle_1\,{=}\, {-}i \langle q \rangle_1$. We have taken $\delta \, {=} \,\omega_m$ and effective detuning $\Delta \,{\equiv}\, \Delta_0 \, {-}\, g \langle q \rangle_0 \,{=} \,\omega_m \, {\gg} \, \kappa$, which means that the frequency component at $\omega_l+\omega_m$ is resonantly enhanced, while the component at $\omega_l-\omega_m$ is significantly suppressed, leading to the fact that $\langle a \rangle_{-1} \,{\ll} \,\langle a \rangle_1$. We can therefore safely neglect this frequency component in the cavity field $\langle a \rangle$. We have also assumed $G_1^2/\kappa\gamma \ll1$ ($G_1\,{=}\,g\langle a \rangle_1$) in deriving $\langle a \rangle_0$, implying that a sufficiently weak ``probe" field is used. The expression of $\langle a \rangle_1 \simeq \frac{E_1}{\kappa C_0}$ for a large cooperativity $C_0\,{=}\,G_0^2/\kappa\gamma \gg 1$ ($G_0\,{=}\,g\langle a \rangle_0$) indicates the OMIT effect: the amplitude of the frequency component $\omega_p \,{=}\, \omega_c$ becomes very small when the red-detuned pump is strong enough.

Apart from nonzero first moments, a coherent state implies that the quantum fluctuations must be as (or very close to) those of the vacuum state. We therefore turn to the quantum dynamics by writing any operator as $O(t)=\langle O \rangle(t) + \delta O(t)$. We assume that $|\langle a \rangle| \gg 1$, allowing us to safely neglect second order terms in the expansion of each $O(t)$. The linearized quantum Langevin equations (QLEs) describing the quantum fluctuations ($\delta q$, $\delta p$, $\delta x$, $\delta y$), with $\delta x=(\delta a + \delta a^{\dag})/\sqrt{2}$ and $\delta y=i(\delta a^{\dag} {-} \delta a)/\sqrt{2}$, are given by
\begin{equation}\label{QLEs}
\begin{split}
\delta \dot{q}&= \omega_m \delta p,   \\
\delta \dot{p}&= -\omega_m \delta q -\gamma \delta p +\!\! \sqrt{2} \sum_{n=0,1} e^{-i n \omega_m t} \big( G_n^x \delta x + G_n^y \delta y \big) +\xi,   \\
\delta \dot{x}&= -\kappa \delta x +\Delta \delta y - \!\! \sqrt{2} \sum_{n=0,1} e^{-i n \omega_m t} G_n^y \delta q  +\!\! \sqrt{2\kappa} x^{\rm in},    \\
\delta \dot{y}&= -\kappa \delta y -\Delta \delta x + \!\! \sqrt{2} \sum_{n=0,1} e^{-i n \omega_m t} G_n^x \delta q  +\!\! \sqrt{2\kappa} y^{\rm in},    \\
\end{split}
\end{equation}
where we have defined $G_n^x\,{=}\,{\rm Re}\,G_n$ and $G_n^y\,{=}\,{\rm Im}\,G_n$ ($n\,{=}\,0,1$), and assumed time-dependent detuning $\tilde{\Delta} \,{=} \, \Delta_0 \, {-}\, g \langle q \rangle \, \, {\simeq} \,\, \Delta$, which is a good approximation when $|E_1| \ll |E_0|$. $\xi$ and $x^{\rm in}$, $y^{\rm in}$ are input noise operators for the mechanical and cavity mode, respectively, which are zero mean and characterized by the correlation functions: $\langle \xi(t)\xi(t')+\xi(t') \xi(t) \rangle/2 \, {\simeq}\,\gamma (2\bar n{+}1) \delta(t{-}t')$ (in the Markovian approximation valid in current experimental regime, and $\bar{n}=\big[ {\rm exp}\big( \frac{\hbar \omega_m}{k_B T} \big) {-}1 \big]^{-1} $ is the mean thermal phonon number), and $\langle x^{\rm in}(t) x^{\rm in}(t')\rangle\,{=}\,\langle y^{\rm in}(t) y^{\rm in}(t')\rangle\,{=}\,\frac{1}{2}\delta(t-t')$. The correlators are different because the way quantum noise affects the field in cavity and the MR. The vacuum noise enters the cavity and thus it gets directly added to the field mode. For the MR the Brownian noise acts on a massive system, i.e., it acts like a force and thus affects the momentum of the MR. The QLEs~\eqref{QLEs} can be conveniently solved in the frequency domain~\cite{SuppM}, and analytical expressions of $\langle \delta q^2 \rangle$ and $\langle \delta p^2 \rangle$ can be achieved, which are, however, too lengthy to be reported here. Nevertheless, we find numerically that the ``probe" field has a negligible effect on the fluctuations $\langle \delta q^2 \rangle$ and $\langle \delta p^2 \rangle$ provided that $|E_1| \,{\ll}\, |E_0|$~\cite{SuppM}. Adopting the parameters from a recent experiment~\cite{Simon17}: $\omega_m/2\pi \,{=}\, 5.25$ GHz, $\gamma \,{=}\, \omega_m/3.8\times 10^5$, $\kappa/2\pi \,{=}\, 846$ MHz, $g/2\pi \,{=} \sqrt{2} \,{\times}\, 869$ KHz, and considering relatively higher temperatures $T\,{=}\,1$ K (10 K), corresponding to $\bar n\,{=}\,3.49$ (39.19), we obtain $\langle \delta q^2 \rangle \,{\simeq}\,\, 0.5139 \,\, (0.5887)+\delta q^2(t)$, and  $\langle \delta p^2 \rangle \,{\simeq}\, 0.5138 \,\, (0.5886)+\delta p^2(t)$ for pump powers $P_0\,{=}\,50$ $\mu$W and $P_1\,{=}\,0.5$ $\mu$W (which gives $|E_1/E_0|\,{=}\,0.1 \,{\ll}\, 1$), where $\delta q^2(t)$ and $\delta p^2(t)$ are fluctuation modulation terms due to the weak ``probe" field. These modulation terms are negligible with respect to $\langle \delta q^2 \rangle \simeq 0.5139$ (0.5887) and $\langle \delta p^2 \rangle \simeq 0.5138$ (0.5886) (corresponding to $\bar{n}_0=0.0138$ (0.0886)) when the ``probe" field is absent~\cite{SuppM}, and thus the MR is prepared approximately in a coherent state. This directly follows from the linearity of Eq.~\eqref{QLEs}.

{\it Adding a phonon onto coherent states.} Once the MR is prepared in the desired coherent state, we switch off the pump and ``probe" fields. After a time $ \kappa^{-1} \, {\ll} \, \tau_{\rm pd} \, {\ll} \, \gamma^{-1}$ (see Fig.~\ref{fig1} (b)), all cavity photons decay and the mechanical state remains effectively unchanged. The system is then in the state $| 0 \rangle_c |\beta \rangle_m$, where $|\beta| \,\, {=} \sqrt{2} |\langle q \rangle_1| \,\,{\simeq} \sqrt{2} \frac{\omega_m  E_1}{\,\,g \, E_0}$, when $\kappa \, {\ll} \, \omega_m$ and $C_0 \, {\gg} \, 1$. Using the parameters of Ref.~\cite{Simon17} and taking $P_0\,{=}\,0.2$ mW, 5.4 pW $\!{<} \, P_1 \,{<} \,48$ pW corresponds to a coherent state with amplitude $1{<} \, |\beta| \,{<}3$, where, as we will show later, the mechanical squeezing is most notable. Note that, in practice, the MR is prepared in a thermal coherent state (e.g., due to the absorption heating) with thermal phonon occupancy $\bar{n}_0 \ll 1$. We shall first consider the case of $\bar{n}_0\,{=}\,0$ and then study the effect of the residual excitations on the results of this idealized case.

Adding a single phonon onto a coherent state can be implemented by sending a {\it weak} blue-detuned write pulse at $\omega_b \simeq \omega_c+\omega_m$, which yields the effective Hamiltonian $H_b=\hbar G_b (a^{\dag} b^{\dag} +a b)$~\cite{SuppM}, where $b\,{=}\,(q\,{+}\,ip)/\!\sqrt{2}$, $G_b\,{=}\,g\sqrt{n_b/2}$ is the effective optomechanical coupling rate, and $n_b\,{=}\,2\kappa P_b/\big[ \hbar \omega_b (\kappa^2{+}\omega_m^2) \big]$ is the intracavity photon number ($P_b$ is the power of the pulse). To simplify the model, we consider flat-top pulses. This Hamiltonian generates a two-mode squeezing interaction  (with a small squeezing parameter since $G_b \tau_b$ is assumed to be small. $\tau_b$ is the pulse duration). The state of the system after the pulse can be approximated as 
\begin{equation}
|\phi \rangle \approx (1+\PP a^{\dag}b^{\dag}) \, | 0 \rangle_c |\beta \rangle_m = | 0 \rangle_c |\beta \rangle_m +\PP | 1 \rangle_c  \big(b^{\dag} |\beta \rangle_m \big), 
\end{equation}
where $|\PP|=G_b \tau_b \ll 1$. The MR is conditionally prepared in a single-PACS $b^{\dag}  |\beta \rangle_m$ if a single photon is detected. In what follows, we derive the exact solution of the system state after applying the write pulse. We consider the pulse duration to be much shorter than the mechanical decoherence time $\tau_b \ll \gamma^{-1}$, such that the decay of mechanical energy can be neglected. This leads to the QLEs during the write pulse: $\delta \dot{a} = -\kappa \delta a + i G_b \delta b^{\dag} +\!\sqrt{2 \kappa} a^{\rm in}$, and $\delta \dot{b} = i G_b \delta a^{\dag}$~\cite{SuppM}. We consider also a weak coupling $G_b \ll \kappa$, allowing for adiabatic elimination of the cavity mode, and we thus have $\delta a \simeq \kappa^{-1} \big( i G_b \delta b^{\dag} +\!\sqrt{2\kappa} a^{\rm in} \big)$. Using the standard input-output formula~\cite{Gardiner}: $a^{\rm out}\,{=}\,\sqrt{2\kappa} \delta a -a^{\rm in} $, we obtain $a^{\rm out} \,{=}\, i\sqrt{2 \GG_b} b^{\dag} + a^{\rm in}$, and $\dot{b} \,{=}\, \GG_b b +i \sqrt{2 \GG_b} a^{\rm in \dag}$, where $\GG_b \,\,\, {\equiv} \,\,\, G_b^2/\kappa$. We introduce the temporal modes for the cavity driven by the write pulse of duration $\tau_b$, $A^{\rm in (out)} (\tau_b) = \big[ {\, \pm}\, 2\GG_b/\,\big( 1 \,{-}\, e^{\mp 2\GG_b \tau_b} \big) \big]^{1/2} \int_0^{\tau_b} e^{\mp \GG_b s} a^{\rm in(out)} (s)\, ds$ ($[A^j, \,\, A^{j \dag}]\,\,\,{=} \,\,\, 1, j \,\,\,{=}\,\,\, \{ {\rm in}, \,\, {\rm out}\} $)~\cite{enOM2}. This leads to the expressions $A^{\rm out} (\tau_b) = e^{\GG_b \tau_b} A^{\rm in} (\tau_b) +i \sqrt{ e^{2 \GG_b \tau_b} -1} b^{\dag}(0)$, and $b (\tau_b) = e^{\GG_b \tau_b} b(0) +i \sqrt{ e^{2 \GG_b \tau_b} -1} A^{\rm in \dag}(\tau_b)$. A propagator $U(\tau_b)$ that satisfies $A^{\rm out} (\tau_b) = U^{\dag} \, A^{\rm in} (\tau_b) \, U$ and $b(\tau_b) = U^{\dag} \, b(0) \, U$ can be extracted, $U(\tau_b) = e^{i \sqrt{ 1-Z(\tau_b)^2 } A^{\rm in \dag} b^{\dag} }  Z(\tau_b)^ {1+ A^{\rm in \dag} A^{\rm in} + b^{\dag} b }  e^{-i \sqrt{1-Z(\tau_b)^2 } A^{\rm in} b }$~\cite{Galland}, where $Z(\tau_b)=e^{-\GG_b \tau_b}$ ($0<Z \le 1$). For an initial state $| 0 \rangle_c |\beta \rangle_m$, the system, at the end of the pulse, is prepared in the state 
\begin{equation}\label{phitau}
\begin{split}
|\phi (\tau_b) \rangle &= Z \, e^{-\frac{|\beta|^2}{2} (1-Z^2) }  \sum_{n=0}^{\infty} \frac{ i^n \big( 1-Z^2 \big)^{\frac{n}{2}} }{n!}   \big( A^{\rm in \dag} b^{\dag} \big)^n   |0\rangle_c  | Z\beta \rangle_m   \\
&\approx  Z \, e^{-\frac{|\beta|^2}{2} (1-Z^2) }  \Big[  |0\rangle_c  | Z\beta \rangle_m \, {+}\, i \sqrt{1{-} Z^2} \, |1\rangle_c \, b^{\dag} | Z\beta \rangle_m  \Big].
\end{split}
\end{equation} 
For taking ``$\approx$" in Eq.~\eqref{phitau}, we have assumed $1-Z(\tau_b)^2 \ll 1$ ($\GG_b \tau_b \ll 1$), i.e., the probability of generating a photon-phonon pair is sufficiently low, such that the possibility of generating more than one photon-phonon pair is negligible. The MR is then prepared with a high probability in a single-PACS $b^{\dag} |Z\beta \rangle_m$, with a slightly reduced amplitude $|\beta| \to |\beta| e^{-\GG_b \tau_b}$, provided that a single photon is detected on cavity resonance. This probability can be as high as $98.8\%$ in the experiment~\cite{Simon17}. This is similar to the proposals~\cite{Galland,Vanner} and experiments~\cite{Simon17,Bellini04} for preparing single photon and single phonon states.

\begin{figure}[b]
\hskip-0.2cm\includegraphics[width=\linewidth]{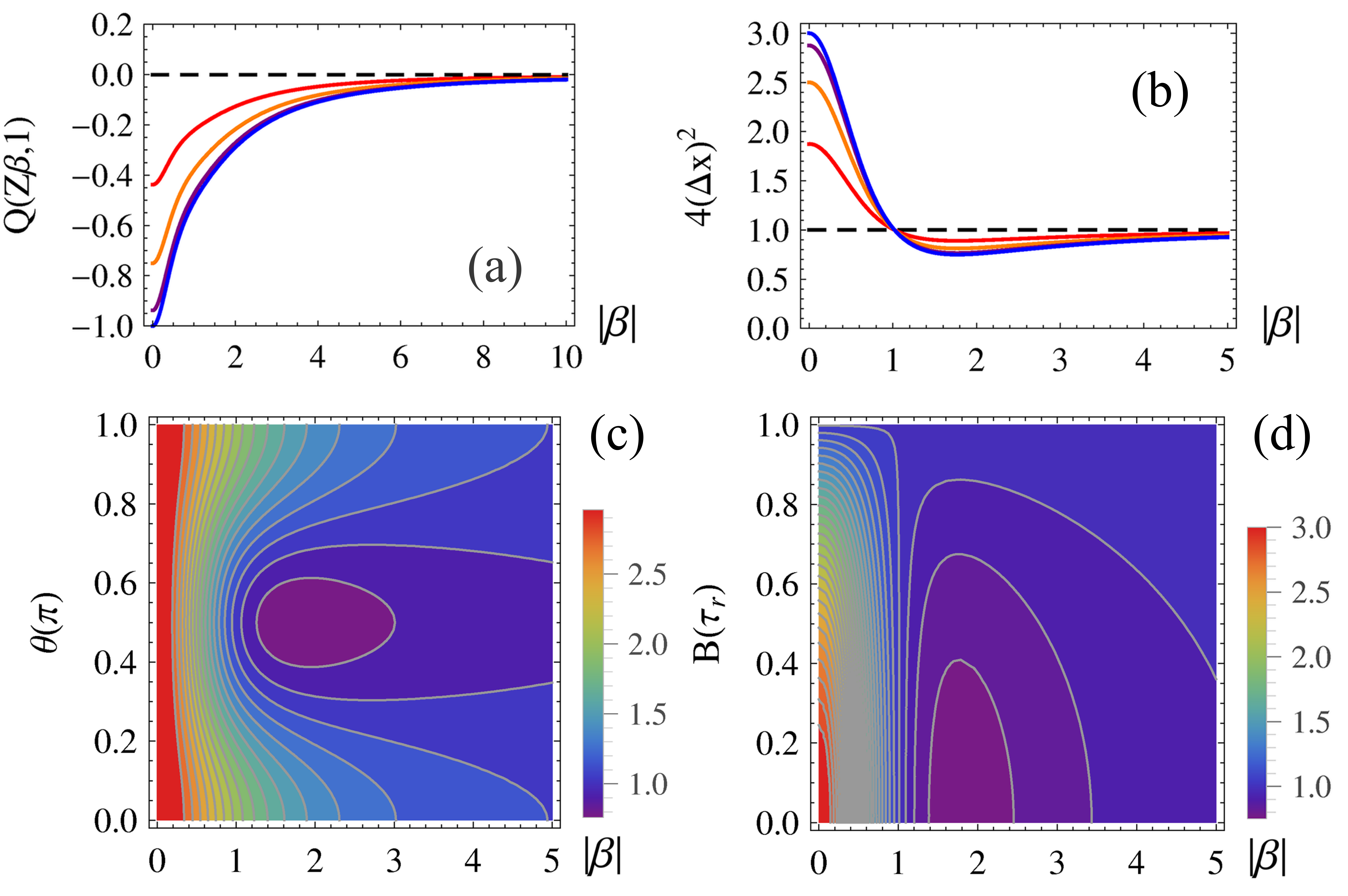} 
\caption{(a) $Q(Z\beta,1)$ \Big[(b) $4(\Delta x_{\frac{\pi}{2}})^2$\Big] vs $|\beta|$ for different values of $B(\tau_r)$: solid lines from bottom to top (top to bottom above the dashed line) correspond to $B=0.01$, 0.25, 0.5, 0.75, respectively. $1{-}B^2$ is the state-swap efficiency of the readout pulse. The dashed line denotes $Q(Z\beta, 0){=}0$ \Big[$4(\Delta x_{\frac{\pi}{2}})^2{=}1$\Big] for coherent states. (c)-(d) Contour plots of $4(\Delta x_{\theta})^2$ versus some key parameters: $B=0.15$ in (c) and $\theta=\frac{\pi}{2}$ in (d). $4(\Delta x_{\theta})^2\,{=}\,1$ corresponds to vacuum fluctuations. We take $\tau_b\, {=}\, \tau_r/4 \,{=}\, 10^{-8}$ s, and $G_b\,{=}\,G_r/5 \,{=}\, 10^8$ Hz $\ll \kappa$ (close to the parameter regime of~\cite{Simon17}), which yield $Z(\tau_b) \simeq 0.98$ and $B(\tau_r) \simeq 0.15$. }
\label{fig2}
\end{figure}

{\it Readout of PACS.} The phonon state can be read out by sending a red-detuned laser pulse at $\omega_r \simeq \omega_c -\omega_m$, which yields the effective Hamiltonian $H_r=\hbar G_r (a^{\dag} b +a b^{\dag} )$~\cite{SuppM}, with the effective coupling rate $G_r=g\sqrt{n_r/2}$, and the intracavity photon number $n_r\,{=}\,2\kappa P_r/\big[ \hbar \omega_r (\kappa^2{+}\omega_m^2) \big]$ ($P_r$ being the power of the pulse). Again, we assume $\tau_r \ll \gamma^{-1}$ and neglect the mechanical damping. The QLEs during the readout pulse are: $\delta \dot{a} = -\kappa \delta a + i G_r \delta b +\!\sqrt{2 \kappa} a^{\rm in}$, and $\delta \dot{b} = i G_r \delta a$~\cite{SuppM}. Following the same procedures as in the last section, we obtain the temporal mode of the cavity output field, $A^{\rm out} (\tau_r) = B(\tau_r) A^{\rm in} (\tau_r) +i \sqrt{1- B(\tau_r)^2 } b(0)$, where $B(\tau_r)=e^{-\GG_r \tau_r}$  ($0<B \le 1$) and $\GG_r \equiv G_r^2/\kappa$. It is clear that $A^{\rm out} (\tau_r) = i b(0)$ when $B(\tau_r) \to 0$, implying that the mechanical state is perfectly transferred to the optical mode (apart from a phase difference). Therefore, the nonclassical features of the generated mechanical state can be directly verified by measuring the properties of the cavity output field.

We calculate the Mandel $Q$ parameter~\cite{Mandel}, $Q(\alpha,m)=\Big[ \langle (A^{\rm out \dag} A^{\rm out} )^2 \rangle -\langle A^{\rm out \dag} A^{\rm out} \rangle^2 \Big] / \langle A^{\rm out \dag} A^{\rm out} \rangle -1$, which is defined for an $m$-PACS $|\alpha, m \rangle \equiv  \NN \, b^{\dag m} |\alpha \rangle$, with the normalization constant $\NN=\big[ m! L_m(-|\alpha|^2) \big]^{-1/2}$, where $ L_m(x)$ is the Laguerre polynomial of order $m$. In the present scheme, we have $m\,\,{=}\,\,1$ and $\alpha \,\,{=}\,\,Z(\tau_b)\beta$. A negative value of $Q(\alpha, m)$ represents the sub-Poissonian character of the state, reflecting its non-Gaussian nature. It is known that Gaussian states correspond to Poissonian or super-Poissonian statistics~\cite{note}. For $m\,\,{=}\,\,0$, $Q(\alpha, 0)\,\,{=}\,\,0$, corresponding to coherent states, while for $\alpha\,\,{=}\,\,0$, $Q(0, m)\,\,{=}\,\,{-}1$. The $Q$ parameter can be calculated more conveniently using the fact that $ \langle (A^{\rm out \dag} A^{\rm out} )^2 \rangle =\langle A^{\rm out \, 2} A^{\rm out \dag \, 2} \rangle \,{-}\, 3 \langle A^{\rm out} A^{\rm out \dag} \rangle \,{+}\, 1$, and $\langle \alpha, m| b^n b^{\dag n}  |\alpha, m \rangle \,{=}\, (n{+}m)! \, L_{n+m} (-|\alpha|^2)/\big[ m! L_m (-|\alpha|^2) \big]$. After straightforward calculations, we obtain $\langle A^{\rm out} A^{\rm out \dag} \rangle=B^2+2(1\,{-}\, B^2)\frac{L_2(-|\alpha|^2)}{L_1(-|\alpha|^2)}$, $\langle A^{\rm out \dag} A^{\rm out} \rangle  = \langle A^{\rm out} A^{\rm out \dag} \rangle -1$, and $\langle A^{\rm out \, 2} A^{\rm out \dag \, 2} \rangle=2B^4 + 8 B^2 (1{-}B^2) \frac{L_2(-|\alpha|^2)}{L_1(-|\alpha|^2)} + 6 (1{-}B^2)^2 \frac{L_3(-|\alpha|^2)}{L_1(-|\alpha|^2)}$, for the state $|0\rangle_c |\alpha, 1\rangle_m$, i.e., the state generated after the write pulse and a single photon being detected (see Fig.~\ref{fig1} (b)). In Fig.~\ref{fig2} (a), we show $Q(Z \beta,1)$ versus $|\beta|$ for different values of $B(\tau_r)$. $Q(Z \beta,1)$ well below zero for small values of $|\beta|$ is a clear sign of the sub-Poissonian character of the state, and a small value of $B(\tau_r)$ is preferred for seeing such a nonclassical feature.

Another important property of the mechanical state $|\alpha, 1\rangle$ is quadrature squeezing~\cite{GA91}, which is a property that a mechanical Fock state~\cite{Galland,Simon17} does not possess. To demonstrate this, we define the quadrature $x_{\theta}$ of the cavity output field, $x_{\theta}=(A^{\rm out} e^{i \theta}+A^{\rm out \dag} e^{-i \theta})/2$, and if its variance $(\Delta x_{\theta})^2=\langle x_{\theta}^2 \rangle - \langle x_{\theta} \rangle^2$ is less than that of the vacuum state, the quadrature $x_{\theta}$ of the field is squeezed, implying that the quadrature of the mechanical mode is squeezed. The expression of $(\Delta x_{\theta})^2$ can be obtained  
\begin{equation}
(\Delta x_{\theta})^2 = \frac{  3{-} 2B^2 {+} (1{-}B^2) \big( e^{i 2\theta} \alpha^2 {+} e^{-i 2\theta}  \alpha^{*2} \big)  {+} 2|\alpha|^2 {+} |\alpha|^4  }{ 4 \big( 1 +|\alpha|^2 \big)^2 } ,
\end{equation}
which becomes $(\Delta x_{\theta})^2=\frac{1}{4}$ for $B=1$, corresponding to the vacuum state of the cavity output. In Fig.~\ref{fig2} (b)-(d), we display $4(\Delta x_{\theta})^2$ versus some key parameters. It shows that $\theta=\frac{\pi}{2}$ and $B(\tau_r)\to 0$ ($B_{\rm min}$ is bounded by $\tau_r \ll \gamma^{-1}$ and $G_r\ll \kappa$) are optimal for observing quadrature squeezing, and considerable squeezing below vacuum has been found in the cavity output field as a result of the nonclassical phonon state.

\begin{figure}[t]
\hskip-0.3cm\includegraphics[width=0.98\linewidth]{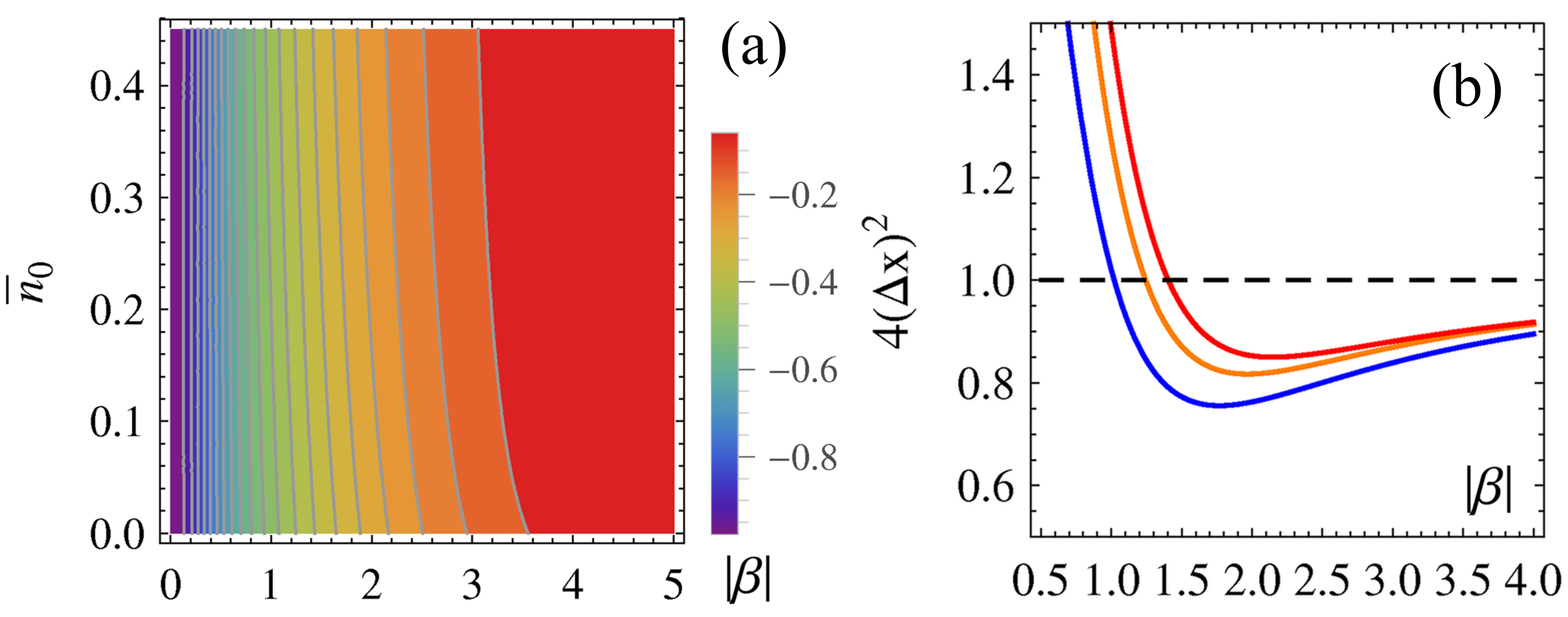}
\caption{(a) Contour plot of the $Q$ parameter versus $|\beta|$ and $\bar{n}_0$. (b) $4(\Delta x_{\frac{\pi}{2}})^2$ versus $|\beta|$ for different values of $\bar{n}_0$: solid lines from bottom to top correspond to $\bar{n}_0=0, 0.2, 0.45$, respectively. We take  $\theta=\frac{\pi}{2}$, $Z(\tau_b)=0.98$, and $B(\tau_r)=0.15$ as in Fig.~\ref{fig2}. }
\label{fig3}
\end{figure}

{\it Effects of residual thermal excitations.} We now discuss the effects of the residual thermal excitations $\bar{n}_0 \ll 1$ in the coherent-state-preparation stage. That is to say, we prepare a thermal coherent state (a thermal state displaced by $|\beta|$ in phase space) with nonzero phonon occupancy~\cite{cool2,Simon17}. Before applying the write pulse, the MR is in the state $\rho_{\rm th,c}\,{=}\,(1{-}s) \sum_{n=0}^{\infty} s^n D(\beta) |n\rangle \langle n| D^{\dag}(\beta)$, with $s\,{=}\,\bar{n}_0/(1{+}\bar{n}_0)$. For $\bar{n}_0 \,{<} \, 0.45$, the MR is most likely (${>}\,90\%$) either in the state $|0\rangle$ or $|1\rangle$. We thus truncate the Fock state basis up to $n\,{=}\,1$, and $\rho_{\rm th,c}$ can be approximated as $\rho_{\rm th,c} \simeq (1{-}s) |\beta \rangle \langle \beta| +(1{-}s) s D(\beta) |1\rangle \langle 1| D^{\dag}(\beta)$. This leads to the conditional state of the MR (unnormalized)~\cite{SuppM} 
\begin{equation}
\begin{split}
\rho_m(\tau_b)\approx \,&b^{\dag} |\alpha \rangle \langle \alpha | b + s \Big( |\beta|^2 b^{\dag} |\alpha \rangle \langle \alpha | b + Z^2  b^{\dag 2} |\alpha \rangle \langle \alpha | b^2  \\
&- \beta^* Z\, b^{\dag} |\alpha \rangle \langle \alpha | b^2 - \beta Z\, b^{\dag 2} |\alpha \rangle \langle \alpha | b  \Big) \\
\end{split}
\end{equation}
after the write pulse and the detection of a single photon, which contains the component of a two-PACS due to the small probability of the initial state in $|1\rangle_m$. By measuring the cavity output of the readout pulse, we obtain the $Q$ parameter, which takes the form $\QQ={\cal N}_Q^{-1} \Big[ \big( 1+s |\beta|^2 \big) Q_1 + s \big( Z^2 Q_2 - \beta^* Z Q_3 -\beta Z Q_4 \big) \Big]$~\cite{SuppM}, where $Q_j = {\rm tr} \Big[ Q |0\rangle_c\langle 0| \otimes b^{\dag j} |\alpha \rangle \langle \alpha | b^j \Big]$ ($j=1,2$), $Q_3 = {\rm tr} \Big[ Q  |0\rangle_c\langle 0| \otimes b^{\dag} |\alpha \rangle \langle \alpha | b^2 \Big]$, and $Q_4 = {\rm tr} \Big[ Q  |0\rangle_c\langle 0| \otimes b^{\dag 2} |\alpha \rangle \langle \alpha | b \Big]$. The normalization constant ${\cal N}_Q=1 + s \big( |\beta|^2 +Z^2 -\beta^* Z -\beta Z \big)$ is introduced to keep $\QQ=0$ (${-}1$) for coherent (Fock) states. Figure~\ref{fig3} (a) shows the effect of the residual excitations on the $Q$ parameter and no appreciable effect is found in particular for small $|\beta|$. A similar form of $(\Delta x_{\theta})^2$ can be derived~\cite{SuppM}, which is more sensitive to the residual excitations, as shown in Fig.~\ref{fig3} (b). Nevertheless, for small $\bar{n}_0$ significant sub-Poissonian character and squeezing of the state are present, characterizing its nonclassical features. We note that for $|\beta|\to 0$, we also generate phonon-added thermal states of the MR. Such states also exhibit nonclassical properties as is known from the corresponding studies on photons~\cite{thermal}.

In conclusion, we have presented a scheme to generate and detect single-PACS of the mechanical motion in optomechanical systems. It is designed for pulses and combines ground state cooling, optomechanical parametric down-conversion and beamsplitter interaction, and single photon detection. Our scheme can also be applied to electromechanical systems and opens promising perspectives for the generation of a series of non-Gaussian states of a mechanical oscillator.  While we discussed phonon addition, the process of subtraction can be carried out by using the red-detuned pump and by detecting a single photon~\cite{Bellini07} for phonons prepared in arbitrary states.

This work has been supported by the Biophotonics initiative of the Texas A{\rm \&}M University and the National Key Research and Development Program of China (Grant No. 2017YFA0304202). We also acknowledge support from Foundation for Fundamental Research on Matter (FOM) Projectruimte grants (15PR3210, 16PR1054), the European Research Council (ERC StG Strong-Q), and the Netherlands Organisation for Scientific Research (NWO/OCW), as part of the Frontiers of Nanoscience program, as well as through a Vidi grant (016.159.369).

\end{document}